\documentclass[aps,prb,twocolumn,groupedaddress,amsmath,eqsecnum]{revtex4}

    \usepackage{bm}
    \usepackage{graphicx}

\newcommand\beq{\begin{equation}}
\newcommand\eeq{\end{equation}}
\newcommand\beqa{\begin{eqnarray}}
\newcommand\eeqa{\end{eqnarray}}
\newcommand{\nn}{\nonumber\\}
\newcommand{\sw}{\text{SW}}
\newcommand{\shs}{\text{SHS}}
\newcommand{\dd}{{d}}
\newcommand{\ee}{{e}}

\begin{document}

\title{How `sticky' are short-range square-well fluids?}

\author{Alexandr Malijevsk\'y}
\email{malijevsky@icpf.cas.cz } \affiliation{E. H\'ala Laboratory of
Thermodynamics, Academy of Science of the Czech Republic, Prague 6,
Czech Republic \\ Institute of Theoretical Physics, Faculty of
Mathematics and Physics, Charles University, Prague 8, Czech
Republic}
\author{Santos B. Yuste}
\email{santos@unex.es}
\homepage{http://www.unex.es/eweb/fisteor/santos/}
\author{Andr\'es Santos}
\email{andres@unex.es}
\homepage{http://www.unex.es/eweb/fisteor/andres/}
\affiliation{Departamento de F\'{\i}sica, Universidad de
Extremadura, E-06071 Badajoz, Spain}
\date{\today}

\begin{abstract}
The aim of this work is to investigate to what extent the structural
properties of a short-range square-well (SW) fluid of range
$\lambda$ at a given packing fraction $\eta$ and reduced temperature
$T^*=k_BT/\epsilon$ can be represented by those of a
sticky-hard-sphere (SHS) fluid at the same packing fraction and an
effective stickiness parameter $\tau(T^*,\lambda)$. Such an
equivalence cannot hold for the radial distribution function $g(r)$
since this function has a delta singularity at contact ($r=\sigma$)
in the SHS case,  while it has a jump discontinuity at
$r=\lambda\sigma$ in the SW case. Therefore, the equivalence is
explored with the cavity function $y(r)$, i.e., we assume that
$y_{\sw}(r|\eta,T^*;\lambda)\approx
y_{\shs}(r|\eta,\tau(T^*,\lambda))$. Optimization of the agreement
between $y_{\sw}$ and $y_{\shs}$ to first order in density suggests
the choice $\tau(T^*,\lambda)=[12(\ee^{1/T^*}-1)(\lambda-1)]^{-1}$.
We have performed Monte Carlo (MC) simulations of the SW fluid for
$\lambda=1.05$, $1.02$, and $1.01$ at several densities and
temperatures $T^*$ such that $\tau(T^*,\lambda)=0.13$, $0.2$, and
$0.5$. The resulting cavity functions  have been compared with MC
data of SHS fluids  obtained by Miller and Frenkel [{J. Phys: Cond.
Matter} \textbf{16}, S4901 (2004)]. Although, at given values of
$\eta$ and $\tau$, some local discrepancies between $y_{\sw}$ and
$y_{\shs}$ exist (especially for $\lambda=1.05$), the SW data
converge smoothly toward the SHS values as $\lambda-1$ decreases. In
fact, precursors of the singularities of $y_{\shs}$ at certain
distances due to geometrical arrangements are clearly observed in
$y_{\sw}$. The approximate mapping $y_{\sw}\to y_{\shs}$ is
exploited to estimate the internal energy and structure factor of
the SW fluid from those of the SHS fluid. Taking for $y_{\shs}$ the
solution of the Percus--Yevick equation  as well as the
rational-function approximation,  the radial distribution function
$g(r)$ of the SW fluid is theoretically estimated and a good
agreement with our MC simulations is found. Finally, a similar study
is carried out for short-range SW fluid mixtures.
\end{abstract}

\maketitle

\section{Introduction\label{sec1}}
It is  well known  that  colloidal particles in a suspension of free
polymers interact through effective attractive  forces (of entropic
origin) with a range  and strength determined by the size and
concentration of the polymers.\cite{HSKGKQ84,KRBDVM89} A simple
 model describing this effective interaction is the square-well
 (SW) potential
\beq
\varphi_{\sw}(r)=\begin{cases}
\infty,& r<\sigma,\\
-\epsilon,& \sigma<r<\lambda\sigma,\\
0,&r>\lambda\sigma,
\end{cases}
\label{1}
\eeq
which accounts for excluded volume effects associated with the
hard-core diameter $\sigma$ plus an attractive layer of relative
width $\lambda-1$ and strength $\epsilon$. Although the SW potential
(with $\lambda \simeq 1.5$) was originally introduced as a simple
model for normal liquids,\cite{BH76} it has become even more useful
in colloidal systems, where one typically has $\lambda\leq 1.1$.

The  interaction parameters $\sigma$ and $\epsilon$ can be used to
define the length and energy units, respectively. Thus, the
concentration of particles can be characterized by the volume
fraction $\eta=\frac{\pi}{6}\rho\sigma^3$, where $\rho$ is the
number density, and the temperature $T$ can be measured in units of
$\epsilon$ as $T^*=k_BT/\epsilon$, where $k_B$ is the Boltzmann
constant. Therefore, at given $\eta$ and $T^*$, only the range
$\lambda$ remains as a free parameter. This degree of freedom
disappears in the so-called sticky-hard-sphere (SHS)
limit,\cite{B68} where one takes the combined limits
$\epsilon\to\infty$ (i.e., $T^*\to 0$) and $\lambda\to 1$, while
keeping constant the stickiness parameter

\beq
\tau^{-1}=12{\ee^{1/T^*}}{(\lambda-1)}.
\label{tau}
\eeq
The parameter $\tau$ plays in the SHS fluid a role equivalent to
that played by the reduced temperature $T^*$ in the SW fluid.
Although, strictly speaking, a monodisperse system of SHS is not
thermodynamically stable,  a small degree of polydispersity is
sufficient to restore stability.\cite{S91} Given its simplicity and
the fact that it can be exactly solved within the Percus--Yevick
(PY) approximation,\cite{B68,PS75} the SHS model has received
considerable attention as a convenient model of colloidal
suspensions.\cite{BT79,SG87,KF88,RR89,JBH91,MMR91,TB93,YS93a,YS93b,RAH95,TKR02,GG02,MF03,GG04,MF04a,MF04b,FGG05,J06}

It is obvious that the thermodynamic and structural properties of
the SW fluid must  approach those of the SHS fluid as $\lambda\to 1$
at fixed $\eta$ and $\tau$. This has been tested, for instance, in
the case of the vapour-liquid critical point.\cite{MF04a} An
interesting problem not yet addressed in detail is  the rate of the
convergence from SW to SHS. In other words, how small must the width
$\lambda-1$  be for the properties of the SW system to be
practically the same as those of an equivalent SHS system?
Comparison between Monte Carlo (MC) simulations\cite{HSKGKQ84} for
an SW fluid with $\lambda=1.1$  and theoretical results for SHS
fluids shows a good agreement
 between the respective structure factors
$S(k)$,\cite{MMR91,YS94,AS01} although a systematic phase shift
between both structure factors exists. This is a reflection of the
fact that the radial distribution function $g(r)$, which is
essentially the inverse Fourier transform of $S(k)$, has important
qualitative differences in both classes of systems. While in the
case of an SW fluid, $g(r)$ is finite at $r=\sigma^+$ and presents a
jump discontinuity at $r=\lambda\sigma$, a delta-peak singularity at
$r=\sigma^+$, followed by other delta and jump singularities at
certain characteristic distances,\cite{SG87,MF04a} appear in the SHS
case.

The aim of this paper is two-fold. First, we want  to assess to what
degree a short-range SW fluid behaves as a suitably chosen
``equivalent'' SHS fluid. To that end, we have performed Monte Carlo
(MC) simulations of SW systems with $\lambda=1.05$, $\lambda=1.02$,
and $\lambda=1.01$, and have compared their structural properties
with recent MC simulations\cite{MF04a} for SHS fluids at the same
packing fraction and effective stickiness. The comparison is not
carried out with the radial distribution function $g(r)$, but with
the much more regular cavity (or background) function
$y(r)=\ee^{\varphi(r)/k_BT}g(r)$. It is observed that, as expected,
the SW cavity functions converge toward the SHS ones as $\lambda$
decreases. However, the results show that the range $\lambda=1.05$
cannot be considered small enough to get a cavity function  hardly
distinguishable from the SHS one. The second point we  address is
the estimate of the structure factor and thermodynamic properties of
short-range SW fluids from the knowledge of the properties of SHS
fluids. As said before, the starting point is the assumption that
both cavity functions are approximately the same. {}The approximate
estimates for the internal energy and structure factor of the SW
systems are tested against actual MC simulations, finding a good
agreement. Also in this context, we take advantage of the exact
solution of the PY equation for SHS,\cite{B68} as well as of a more
refined theory,\cite{YS93b} to propose  analytic approximations for
the radial distribution function of SW fluids. A parallel study is
carried out for short-range SW fluid mixtures. Due to the scarcity
of simulation results for SHS mixtures,  we compare our MC results
with the solution of the PY theory.\cite{PS75} It is observed that
in the case of mixtures the influence of $\lambda\neq 1$ on the
mapping SW$\to$SHS is similar to that in the one-component case.

This paper is organized as follows. The basic equations for SW and
SHS fluids are summarized in Section \ref{sec1bis}. The criterion
followed in this paper to define the effective stickiness parameter
and   some  consequences are elaborated in Section \ref{sec2}.
Section \ref{sec3} presents the comparison between our MC
simulations for SW and those reported in Ref.\ \onlinecite{MF04a}
for SHS, while the comparison with theoretical predictions is
presented in Section \ref{sec4}. The case of SW mixtures is
considered in Section \ref{sec5}. Finally, the paper ends with some
concluding remarks.

\section{Basic equations\label{sec1bis}}
In this Section we introduce the notation and display the basic
equations that will be needed in the paper.
\subsection{The square-well fluid}
The SW interaction potential is given by Eq.\ (\ref{1}). Henceforth,
the distance $r$ is assumed to be measured in units of the hard-core
diameter $\sigma$, so that we take $\sigma=1$. The relevant physical
information about the system is contained in the radial distribution
function $g_{\sw}(r|\eta,T^*;\lambda)$, where the notation
emphasizes that it depends on the thermodynamic state (characterized
by the packing fraction $\eta$ and the reduced temperature $T^*$)
and on the relative range $\lambda$. The corresponding cavity
function is
$y_{\sw}(r|\eta,T^*;\lambda)=\ee^{\varphi_\sw(r)/k_BT}g_{\sw}(r|\eta,T^*;\lambda)$.
Inverting this relationship, one has
\beq
g_{\sw}(r|\eta,T^*;\lambda)=
\begin{cases}
0,&r<1,\\
\ee^{1/T^*}y_{\sw}(r|\eta,T^*;\lambda),&1< r <\lambda,\\
y_{\sw}(r|\eta,T^*;\lambda),&r>\lambda.
\end{cases}
\label{7}
\eeq
Since the cavity function is continuous everywhere,\cite{HM86} Eq.\
(\ref{7}) implies that the radial distribution is discontinuous not
only at $r=1$, but also at $r=\lambda$, namely
$g_{\sw}(\lambda^-|\eta,T^*;\lambda)/g_{\sw}(\lambda^+|\eta,T^*;\lambda)=\ee^{1/T^*}$.

The structure factor is defined as $S(k)=1+\rho\widetilde{h}(k)$,
where $\widetilde{h}(k)$ is the Fourier transform of the total
correlation function $h(r)\equiv g(r)-1$. In the case of the SW
fluid,  $S(k)$ can be obtained from $y(r)$ as
\begin{widetext}
\beqa
S_{\sw}(k|\eta,T^*;\lambda)&=&1+\frac{24\eta}{k}\int_0^\infty \dd
r\, r \sin kr\left[g_{\sw}(r|\eta,T^*;\lambda)-1\right]\nn
&=&1+\frac{24\eta}{k}\left\{-\int_0^1\dd r\, r{\sin kr}+
(\ee^{1/T^*}-1)  \int_1^\lambda \dd r\, r{\sin
kr}\,y_{\sw}(r|\eta,T^*;\lambda) \right.\nn && \left.+ \int_1^\infty
\dd r\, r{\sin
kr}\left[y_{\sw}(r|\eta,T^*;\lambda)-1\right]\right\}.
\label{15}
\eeqa

The thermodynamic properties can also be expressed in terms of the
cavity function.
 The virial equation yields
\beqa
Z_{\sw}(\eta,T^*;\lambda)&=&1-\frac{4\eta}{k_BT}\int_0^\infty \dd
r\, r^3 g_{\sw}(r|\eta,T^*;\lambda)\frac{d}{dr}\varphi_\sw(r)\nn
&=&1+4\eta\left[\ee^{1/T^*}y_{\sw}(1|\eta,T^*;\lambda)-(\ee^{1/T^*}-1)\lambda^3y_{\sw}(\lambda|\eta,T^*;\lambda)
\right],
\label{ZSW}
\eeqa
where  $Z\equiv p/\rho k_BT$ is the compressibility factor, $p$
being the pressure. The compressibility equation gives
\beqa
\chi_{\sw}(\eta,T^*;\lambda)&=&S_{\sw}(0|\eta,T^*;\lambda)\nn
&=&1+{24\eta}\left\{ (\ee^{1/T^*}-1)\int_1^\lambda \dd r\,
r^2y_{\sw}(r|\eta,T^*;\lambda) + \int_1^\infty \dd r\,
r^2\left[y_{\sw}(r|\eta,T^*;\lambda)-1\right]-\frac{1}{3}\right\},
\label{15bis}
\eeqa
\end{widetext}
where $\chi\equiv k_BT (\partial \rho/\partial p)_{T}$ is the
isothermal susceptibility. Finally, the excess internal energy per
particle $u$  is
\beqa
\label{uex1}
   \frac{1}{\epsilon}u_{\sw}(\eta,T^*;\lambda)&=&\frac{\rho}{2\epsilon}\int \dd\mathbf{r}\, \varphi_{\sw}(r)g_{\sw}(r|\eta,T^*;\lambda)\nn
    &=&-12 \eta  \ee^{1/T^*} \int_1^\lambda \dd r\, r^2
    y_{\sw}(r|\eta,T^*;\lambda).\nn
 \eeqa
Therefore, $-2u_\sw/\epsilon$ coincides with  the coordination
number.

\subsection{The sticky-hard-sphere limit}
The SHS system can be obtained from the SW one by formally taking
the combined limits $\lambda\to 1$ and $T^*\to 0$ with  the
stickiness parameter (\ref{tau}) kept constant:
\beq
y_{\shs}(r|\eta,\tau)=\lim_{T^*\to 0,\lambda\to
1}y_{\sw}(r|\eta,T^*;\lambda).
\label{2}
\eeq
Equation (\ref{7}) then reduces to
\beq
g_{\shs}(r|\eta,\tau)=\left[\frac{1}{12\tau}\delta(r-1)+\Theta(r-1)\right]y_{\shs}(r|\eta,\tau),
\label{8}
\eeq
so that  the radial distribution function becomes identical to the
cavity function for $r>1$, but a delta-peak of amplitude
proportional to the stickiness appears at contact.

In the SHS limit, Eqs.\ (\ref{15})--(\ref{uex1}) become
\begin{widetext}
\beq
S_{\shs}(k|\eta,\tau)=1+\frac{24\eta}{k}\left\{-\int_0^1\dd r\,
r{\sin kr}+ \frac{\sin k}{12\tau} y_{\shs}(1|\eta,\tau)+
\int_1^\infty \dd r\, r{\sin
kr}\left[y_{\shs}(r|\eta,\tau)-1\right]\right\},
\label{16}
\eeq
\beq
Z_{\shs}(\eta,\tau)=1+4\eta\left[\left(1-\frac{1}{4\tau}\right)y_{\shs}(1|\eta,\tau)-\frac{1}{12\tau}y'_{\shs}(1|\eta,\tau)
\right]
\label{ZShs}
\eeq
\beq
\chi_{\shs}(\eta,\tau)=1+{24\eta}\left\{\frac{1}{12\tau}
y_{\shs}(1|\eta,\tau)+ \int_1^\infty \dd r\,
r^2\left[y_{\shs}(r|\eta,\tau)-1\right]-\frac{1}{3}\right\},
\label{16bis}
\eeq
\end{widetext}
\beq
\label{uex2}
    \frac{1}{\epsilon}u_{\shs}(\eta,\tau)=-\frac{\eta}{\tau}y_{\shs}(1|\eta,\tau).
 \eeq
In Eq.\ (\ref{ZShs}), $y'(r)$ denotes the first derivative of
$y(r)$.

\section{Effective ``stickiness''\label{sec2}}
Now we assume an SW fluid with a small, but non-zero, value of the
well width $\lambda-1$. It is natural to expect that its structural
properties are, at least to some extent, close to those  of an
 SHS fluid characterized by suitably chosen  effective packing fraction
$\eta_{\text{eff}}$ and effective stickiness
$\tau_{\text{eff}}^{-1}$. Comparison between Eqs.\ (\ref{7}) and
(\ref{8}) shows that the radial distribution function $g(r)$ is not
the adequate quantity to establish the approximate mapping
SW$\to$SHS. This also applies to the structure factor $S(k)$ since
it is essentially the Fourier counterpart of  $g(r)$. Therefore, we
choose the cavity function $y(r)$, which is a more regular function
than $g(r)$. Thus, we consider here the approximation
\beq
y_{\sw}(r|\eta,T^*;\lambda)\approx
y_{\shs}(r|\eta_{\text{eff}}(\eta,T^*;\lambda),\tau_{\text{eff}}(\eta,T^*;\lambda)).
\label{3}
\eeq
Note that this implies $g_{\sw}(r|\eta,T^*;\lambda)\neq
g_{\shs}(r|\eta_{\text{eff}}(\eta,T^*;\lambda),\tau_{\text{eff}}(\eta,T^*;\lambda))$
and $S_{\sw}(k|\eta,T^*;\lambda)\neq
S_{\shs}(k|\eta_{\text{eff}}(\eta,T^*;\lambda),\tau_{\text{eff}}(\eta,T^*;\lambda))$.
The next step consists of defining criteria to determine the
effective quantities $\eta_{\text{eff}}(\eta,T^*;\lambda)$ and
$\tau_{\text{eff}}(\eta,T^*;\lambda))$. There exist some  obvious
constraints. First, one must recover the hard-sphere (HS) cavity
function from  $y_{\sw}$  if either $\lambda\to 1$ at fixed $T^*$ or
$T^*\to\infty$ at fixed $\lambda$. More explicitly,
\beq
\lim_{\lambda\to
1}y_{\sw}(r|\eta,T^*;\lambda)=y_{\text{HS}}(r|\eta),
\eeq
\beq
\lim_{T^*\to\infty}y_{\sw}(r|\eta,T^*;\lambda)=y_{\text{HS}}(r|\eta),
\eeq
\beq
\lim_{\tau\to\infty}y_{\shs}(r|\eta,\tau)=y_{\text{HS}}(r|\eta).
\eeq
This implies that
\beq
\lim_{\lambda\to 1} \eta_{\text{eff}}(\eta,T^*;\lambda)=\eta,\quad
\lim_{\lambda\to 1}\tau_{\text{eff}}(\eta,T^*;\lambda)=\infty,
\label{4.1}
\eeq
\beq
\lim_{T^*\to\infty} \eta_{\text{eff}}(\eta,T^*;\lambda)=\eta,\quad
\lim_{T^*\to\infty}\tau_{\text{eff}}(\eta,T^*;\lambda)=\infty.
\label{4.2}
\eeq
Moreover,  Eq.\ (\ref{3}) must become exact if the SHS limit is
taken on the left-hand side and Eq.\ (\ref{2}) is applied. Thus,
\beq
\lim_{T^*\to 0,\lambda\to
1}12(\lambda-1)\ee^{1/T^*}\tau_{\text{eff}}(\eta,T^*;\lambda)=1.
\label{4}
\eeq

As is usual,\cite{KRBDVM89,SG87,RR89,MMR91,TB93,MF04a,VL00} let us
assume for simplicity that
$\eta_{\text{eff}}(\eta,T^*;\lambda)=\eta_{\text{eff}}(\eta,\lambda)$
is independent of temperature and
$\tau_{\text{eff}}(\eta,T^*;\lambda)=\tau_{\text{eff}}(T^*;\lambda)$
is independent of density. The first choice allows us to determine
$\eta_{\text{eff}}$ by applying the first equality of (\ref{4.2}),
which holds regardless of the value of $\lambda \geq 1$. It is then
obvious that
\beq
\eta_{\text{eff}}(\eta,\lambda)=\eta.
\label{5}
\eeq
On the other hand, the choice $\eta_{\text{eff}}(\eta,\lambda)=\eta
\lambda^3$ adopted by Menon et al.\cite{MMR91} is in conflict with
the HS limit (\ref{4.2}).

In order to determine $\tau_{\text{eff}}(T^*,\lambda)$, several
criteria can be used. The conventional one\cite{SG87,RR89,TB93,VL00}
is to impose that the second virial coefficient of the SW fluid be
the same as that of the SHS fluid, i.e.,
$B_2^{\sw}(T^*;\lambda)=B_2^{\shs}(\tau_{\text{eff}}(T^*,\lambda))$.
This yields
\beq
\tau_{\text{eff}}(T^*,\lambda)=\frac{1}{12(\ee^{1/T^*}-1)\alpha(\lambda)}
\label{6}
\eeq
with $\alpha(\lambda)=\alpha_3(\lambda)$, where
\beq
\alpha_n(\lambda)\equiv\frac{1}{n}(\lambda^n-1).
\label{12}
\eeq
The form (\ref{6}) with $\alpha(\lambda)=\alpha_n(\lambda)$ is
consistent with the constraints (\ref{4.1})--(\ref{4}), regardless
of the value of $n$. Since in the second virial coefficient one
simply makes use of the zero-density cavity function $y(r)\to 1$,
the choice $\alpha(\lambda)=\alpha_3(\lambda)$ is not necessarily
the most efficient one from the point of view of Eq.\ (\ref{3}) at
finite density. Another common choice\cite{MMR91,MF04a} is
$\tau_{\text{eff}}(T^*,\lambda)=\ee^{-1/T^*}/12(1-\lambda^{-1})$,
but it is inconsistent with the HS constraint (\ref{4.2}). The same
happens with the direct extrapolation of (\ref{tau}) at finite
$T^*$, i.e.,
$\tau_{\text{eff}}(T^*,\lambda)=\ee^{-1/T^*}/12(\lambda-1)$.

\begin{figure}
\includegraphics[width=\columnwidth]{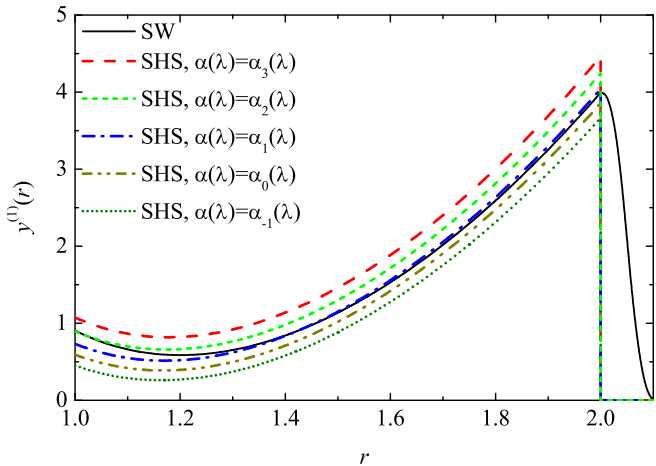} \caption{(Color online) Plot of the first-order
virial coefficient of the cavity function, $y^{\text{(1)}}(r)$, for
an SW fluid with $\lambda=1.05$ and $T^*=0.35$ and for SHS fluids
with values of $\tau$ given by Eqs.\ (\protect\ref{6}) and
(\protect\ref{12}) with $n=3$, $n=2$, $n=1$, $n\to 0$, and $n=-1$.}
\label{fig1}
\end{figure}

As a guide to choose $\tau_{\text{eff}}(T^*,\lambda)$, let us resort
to the exact cavity function to first order in density.\cite{BH67}
In that case one has
\beq
y_{\sw}(r|\eta,T^*;\lambda)=1+y_{\sw}^{(1)}(r|T^*;\lambda)\eta+\mathcal{O}(\eta^2),
\label{A1}
\eeq
\beq
y_{\shs}(r|\eta,\tau)=1+y_{\shs}^{(1)}(r|\tau)\eta+\mathcal{O}(\eta^2),
\label{A6}
\eeq
where the expressions of $y_{\sw}^{(1)}(r|T^*;\lambda)$ and
$y_{\shs}^{(1)}(r|\tau)$ are given in the Appendix. Comparison
between both functions suggests to keep Eq.\ (\ref{6}), except that
it remains to determine $\alpha(\lambda)$ subject to the constraint
$\lim_{\lambda\to 1}\alpha(\lambda)/(\lambda-1)= 1$. In the region
$1<r<2$ we have
\beqa
y_{\sw}^{(1)}(r|T^*;\lambda)&=&8-6r+\frac{1}{2}r^3\nn
&&+\frac{3(\lambda^2-1)^2
r^{-1}-8(\lambda^3-1)+6(\lambda^2-1)r}{12\tau_{\text{eff}}(T^*,\lambda)\alpha(\lambda)}\nn
&&+ \frac{3(\lambda^2-1)^2
r^{-1}}{[12\tau_{\text{eff}}(T^*,\lambda)\alpha(\lambda)]^2},
\label{10}
\eeqa
\beq
y_{\shs}^{(1)}(r|\tau)=8-6r+\frac{1}{2}r^3+\frac{r-2}{\tau}+ \frac{
r^{-1}}{12\tau^2}.
\label{11}
\eeq
It is clear that no unique choice of $\alpha(\lambda)$ can make
$y_{\sw}^{(1)}(r|T^*;\lambda)=y_{\shs}^{(1)}(r|\tau_{\text{eff}}(T^*,\lambda))$
for arbitrary $r$, $T^*$, and $\lambda$. In the  limit of very low
temperatures ($T^*\to 0\Rightarrow \tau_{\text{eff}}\to 0$) the
respective last terms on the right-hand sides of Eqs.\ (\ref{10})
and (\ref{11}) dominate, so that $y_{\sw}^{(1)}(r|T^*;\lambda)\to
y_{\shs}^{(1)}(r|\tau_{\text{eff}}(T^*,\lambda))$ provided that one
takes $\alpha(\lambda)=\alpha_2(\lambda)$, where $\alpha_n(\lambda)$
is defined by Eq.\ (\ref{12}). Nevertheless, for temperatures such
that $12\tau_{\text{eff}}\sim 1$, all the terms in Eqs.\ (\ref{10})
and (\ref{11}) are of the same order, so that again
$\alpha(\lambda)=\alpha_2(\lambda)$ might not be the best choice.
Restricting ourselves, for the sake of simplicity,  to  functions
$\alpha(\lambda)$ of the form (\ref{12}) with $n=\text{integer}$, it
turns out that the optimal choice is generally
$\alpha(\lambda)=\alpha_1(\lambda)$ for temperatures close to the
critical one, i.e., if $\tau_{\text{eff}}\approx 0.1$.\cite{MF04b}
This is illustrated in Fig.\ \ref{fig1} for the case $\lambda=1.05$
and $T^*=0.35$. While the choice
$\alpha(\lambda)=\alpha_2(\lambda)=\frac{1}{2}(\lambda^2-1)$ gives a
value of $y_{\shs}^{(1)}$ in excellent agreement with
$y_{\sw}^{(1)}$ at $r=1$, the best global agreement is provided by
the choice $\alpha(\lambda)=\alpha_1(\lambda)=\lambda-1$. For larger
temperatures the optimal choice could change to
$\alpha(\lambda)=\alpha_0(\lambda)=\ln\lambda$ or even
$\alpha(\lambda)=\alpha_{-1}(\lambda)=1-\lambda^{-1}$. Since we want
to keep $\alpha(\lambda)$ independent of temperature, in what
follows we adopt the choice $\alpha(\lambda)=\alpha_{1}(\lambda)$
for simplicity and also to enhance the cases with
$\tau_{\text{eff}}\approx 0.1$. In summary, the definition of the
effective stickiness parameter is taken as
\beq
\tau_{\text{eff}}(T^*,\lambda)=\frac{1}{12(\ee^{1/T^*}-1)(\lambda-1)}.
\label{13}
\eeq
This is the same choice as considered by de~Kruif et
al.\cite{KRBDVM89}. With this definition, the difference between
$y_{\sw}^{(1)}$ and $y_{\shs}^{(1)}$ in the range $1<r<2$ becomes
\begin{widetext}
\beq
y_{\sw}^{(1)}(r|T^*,\lambda)-y_{\shs}^{(1)}(r|\tau_{\text{eff}}(T^*,\lambda))=\frac{1+6(2-4r+r^2)\tau_{\text{eff}}}{12
r\tau_{\text{eff}}^2} (\lambda-1)+\mathcal{O}((\lambda-1)^2).
\label{18}
\eeq
\end{widetext}

Let us now explore the consequences of the approximation (\ref{3}),
complemented by Eqs.\  (\ref{5}) and (\ref{13}), on some of the
physical properties. We first note that a simple approximate
expression can be obtained for the excess internal energy of the SW
fluid in terms of the value of the SHS cavity function and its slope
at $r=1$. First, we make the approximation $y_{\sw}(r)\approx
y_{\shs}(1)+y_{\shs}'(1) (r-1)$ inside the integral of Eq.\
(\ref{uex1}). Next we perform the integral, make use of Eq.\
(\ref{13}) and neglect terms nonlinear in $\lambda-1$. The result is
\beq
-\frac{1}{\epsilon} u_{\sw}\approx -\frac{1}{\epsilon} u_{\shs}
\left[1+\left(1+12\tau_{\text{eff}}+
\frac{y_{\shs}'(1)}{2y_{\shs}(1)}\right)(\lambda-1)\right],
\label{14}
\eeq
where the arguments denoting the dependence of the quantities on
$\eta$, $T^*$ and $\lambda$ have been omitted for simplicity.
Equation (\ref{14}) shows that, even if the approximation (\ref{3})
is satisfied, $u_{\sw}(\eta,T^*;\lambda)\neq
u_{\shs}(\eta,\tau_{\text{eff}}(T^*;\lambda))$.

Analogously, making  $y_{\sw}(r)\approx y_{\shs}(1)+y_{\shs}'(1)
(r-1)$ in the interval $1<r<\lambda$, Eq.\ (\ref{15}) yields
\beqa
S_{\sw}(k)- S_{\shs}(k)&\approx&\frac{\eta}{k\tau_{\text{eff}}
}\left[y_{\shs}(1)(k\cos k+\sin k)\right.\nn
&&\left.+y_{\shs}'(1)\sin k\right](\lambda-1),
\label{17}
\eeqa
where again terms nonlinear in $\lambda-1$ have been neglected and
the arguments denoting the dependence of the quantities on $\eta$,
$T^*$, and $\lambda$ have been omitted. Taking the limit $k\to 0$ in
the above expression, we have
\beq
\chi_{\sw}- \chi_{\shs}\approx\frac{\eta}{\tau_{\text{eff}}
}\left[2y_{\shs}(1)+y_{\shs}'(1)\right](\lambda-1).
\label{17bis}
\eeq
Finally, from Eqs.\ (\ref{ZSW}) and (\ref{ZShs}) one gets
\beq
Z_{\sw}-
Z_{\shs}\approx-\frac{\eta}{\tau_{\text{eff}}}\left[y_{\shs}(1)+\frac{1}{6}y_{\shs}''(1)\right](\lambda-1)
\label{ZSWSHS}
\eeq

Equations (\ref{14})--(\ref{ZSWSHS}) must be interpreted as simple
heuristic approximations relating some physical properties of SHS
and short-range SW fluids. Thus, they do not give the first-order
terms in a systematic expansion in powers of $\lambda-1$ because
they are based on the ansatz (\ref{3}), which ignores terms of order
$\lambda-1$ in the difference $y_{\sw}(r)- y_{\shs}(r)$, as shown by
Eq.\ (\ref{18}). Note also that Eqs.\ (\ref{14})--(\ref{ZSWSHS})
depend on our choice (\ref{13}) for $\tau_{\text{eff}}$.

\section{Comparison with simulations of sticky hard spheres\label{sec3}}
We have performed conventional Monte Carlo simulations using the
canonical ensemble and employing the linked cell-list method in a
cubic box with a standard periodic boundary conditions. $N = 4000$
particles interacting via the SW potential (\ref{1}) were displaced
according to the Metropolis algorithm to create a sample of
configurations. The calculation of the desired quantities (internal
energy and radial distribution function) was organized in cycles.
Any  randomly chosen particle was attempted to move and the cycle
was repeated $100\times N$ times for equilibration before the system
was analyzed. The process was repeated $1000$ times to get one
averaged set of values. Each run was divided into 20 such blocks, so
that  at least $10^9$ configurations were generated. The distance of
attempted displacements was a mixture of one value adjusted for each
$\lambda$ so that the acceptance ratio was  10\%--15\% and another
value ten times shorter to take care of tiny attractive effects.

\begin{table}
\begin{ruledtabular}
\begin{tabular}{ccccccccc}
   $\eta$
  & $\tau_{\text{eff}}$ & $\lambda$
         & $T^*$
         & $y(1)$ &$y'(1)$ & $-u/\epsilon$ &Eq.\ (\protect\ref{14})\\
\hline
$0.164$&$0.13$&$1.05$&$0.381$&$0.904 $&$-0.314$&$1.271 $&$1.273$\\
&&$1.02$&$0.286$&$0.907 $&$-0.293$&$1.203 $&$1.191$\\
&&$1.01$&$0.239$&$0.908 $&$-0.307$&$1.185 $&$1.164$\\
&&$1$\footnotemark[1]&$0$\footnotemark[1]&$0.901 $\footnotemark[1]&$-0.280 $\footnotemark[1]&$1.137$\footnotemark[1]&$1.137$\footnotemark[1]\\
$0.32$&$0.2$&$1.05$&$0.448$&$0.935 $&$-0.757$&$1.724 $&$1.709$\\
&&$1.02$&$0.324$&$0.927 $&$-0.685$&$ 1.573$&$1.573$\\
&&$1.01$&$0.266$&$0.924 $&$-0.708$&$1.520 $&$1.528$\\
&&$1$\footnotemark[1]&$0$\footnotemark[1]&$0.927 $\footnotemark[1]&$-0.661 $\footnotemark[1]&$1.483 $\footnotemark[1]&$1.483 $\footnotemark[1]\\
$0.4$&$0.5$&$1.05$&$0.682$&$1.622 $&$-4.161$&$1.664 $&$1.604$\\
&&$1.02$&$0.448$&$1.580$&$-3.773$&$1.408 $&$1.387$\\
&&$1.01$&$0.348$&$1.563 $&$-3.697$&$1.328 $&$1.314$\\
&&$1$\footnotemark[1]&$0$\footnotemark[1]&$1.552
$\footnotemark[1]&$-3.606 $\footnotemark[1]&$1.242
$\footnotemark[1]&$1.242 $\footnotemark[1] \footnotetext[1]{The
cases with $\lambda=1$ correspond to MC simulations for SHS
performed by Miller and Frenkel.\protect\cite{MF04a}}
   \end{tabular}
 \caption{SW systems simulated and
corresponding values of $y(1)$, $y'(1)$, and $-u/\epsilon$.}
\label{table1}
\end{ruledtabular}
\end{table}

\begin{figure}
\includegraphics[width=\columnwidth]{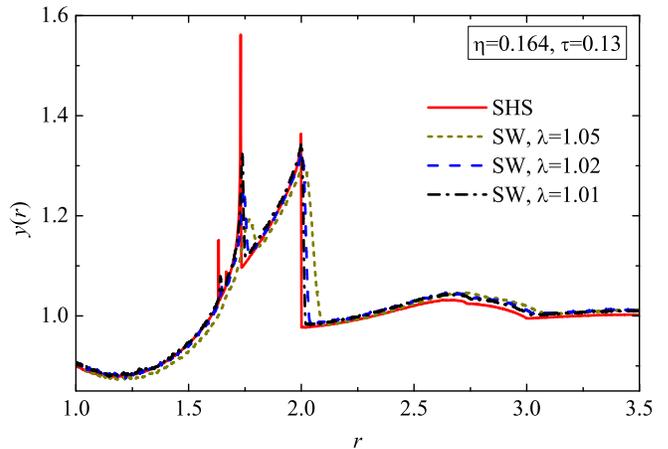} \caption{(Color online) Cavity function at
$\eta=0.164$ and $\tau_{\text{eff}}=0.13$. The solid line
corresponds to MC simulations for SHS by Miller and
Frenkel.\protect\cite{MF04a} The dotted, dashed, and dash-dotted
curves correspond to our simulations for SW systems with
$\lambda=1.05$, $1.02$, and $1.01$, respectively.}
\label{fig2}
\end{figure}

\begin{figure}
\includegraphics[width=\columnwidth]{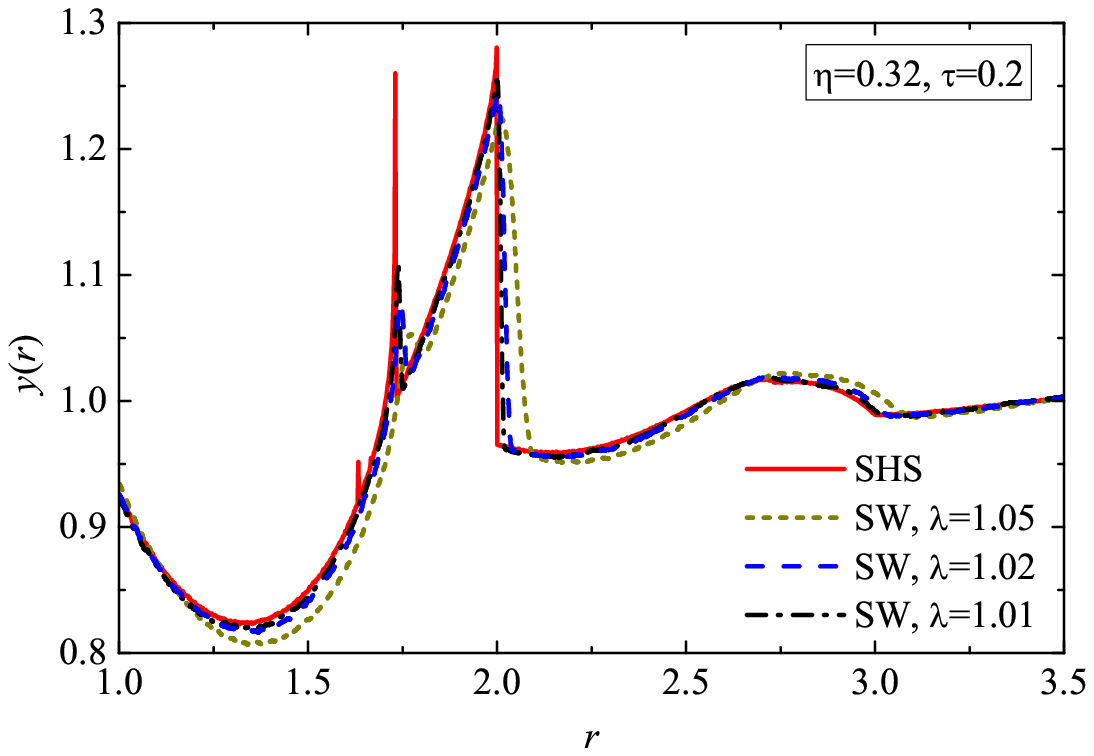} \caption{(Color online) Same as in Fig.\
\protect\ref{fig1} but for $\eta=0.32$ and $\tau_{\text{eff}}=0.2$.}
\label{fig3}
\end{figure}

\begin{figure}
\includegraphics[width=\columnwidth]{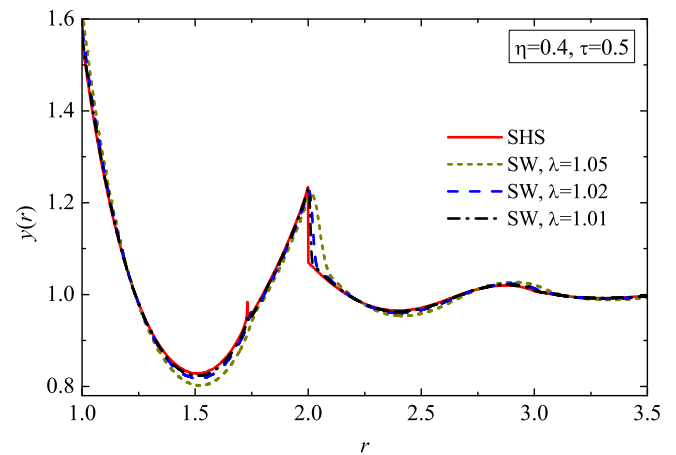} \caption{(Color online) Same as in Fig.\
\protect\ref{fig1} but for $\eta=0.4$ and $\tau_{\text{eff}}=0.5$.}
\label{fig4}
\end{figure}

We have considered three different ranges: $\lambda=1.05$, $1.02$.
and $1.01$. In each case, a temperature $T^*$ has been chosen such
that the effective stickiness parameter $\tau_{\text{eff}}$ defined
by Eq.\ (\ref{13}) takes the same value. In order to test the
approximation (\ref{3}), the values of  $\eta$ and
$\tau_{\text{eff}}$  have been taken  the same as those considered
by Miller and Frenkel in Ref.\ \onlinecite{MF04a}. They are listed
in Table \ref{table1}, where also the internal energy $-u/\epsilon$
and the values at contact  $y(1)$ and $y'(1)$ are included. The two
latter quantities have been obtained from a quadratic fit of the
simulation data for $y_\sw(r)$  and $y_\shs(r)$ in the interval
$1<r<1.1$. Inasmuch as Eq.\ (\ref{3}) is a reasonable approximation,
one would expect that, at a given state point
$(\eta,\tau_{\text{eff}})$, the four cavity functions, i.e.,
$y_{\shs}(r)$ and $y_{\sw}(r)$ with $\lambda=1.05$, $1.02$, and
$1.01$, overlap to some extent. Table \ref{table1} shows that the
contact value of the cavity function, $y(1)$, and its first
derivative, $y'(1)$, are indeed quite similar for the four systems
at each state. At the smallest temperature
($\tau_{\text{eff}}=0.13$), the values of $y(1)$ agree within
statistical errors. At the intermediate temperature
($\tau_{\text{eff}}=0.2$), $y_{\sw}(1)$ for $\lambda=1.05$ deviates
from $y_{\shs}(1)$ less than 1\%, while $y_{\sw}(1)$ for
$\lambda=1.02$ and for $\lambda=1.01$ practically coincide with
$y_{\shs}(1)$. At the highest temperature ($\tau_{\text{eff}}=0.5$),
$y_{\sw}(1)$ differs from $y_{\shs}(1)$ less than 5\%, 2\%, and 1\%
for $\lambda=1.05$, 1.02, and 1.01, respectively. The fact that the
agreement between $y_{\sw}(1)$ and $y_{\shs}(1)$ slightly decreases
as the temperature increases is associated with the choice of
$\tau_{\text{eff}}$ given by Eq.\ (\ref{13}). As discussed in Sec.\
\ref{sec2}, that choice was expected to be better for temperatures
such that $\tau_{\text{eff}}\sim 0.1$ than for larger ones. Similar
conclusions can be drawn from the comparison of $y_{\sw}'(1)$ and
$y_{\shs}'(1)$, although in that case we have observed that the
influence of the noise of the simulation data and  that of the
fitting procedure are slightly larger than in the case of $y(1)$.

Even if $y_{\sw}(r)\approx y_{\shs}(r)$, that does not mean that
$u_{\sw}\approx u_{\shs}$ since the excess internal energy monitors
the correlation function in the range $1<r<\lambda$ and so it is
sensitive to the width of the well. In fact, Table \ref{table1}
shows that the magnitude of the excess internal energy changes  much
more than the contact value of the cavity function when shrinking
the well width. On the other hand, Eq.\ (\ref{14}), which is
implemented in Table \ref{table1} with the simulation data of
$y_\shs(1)$ and $y_\shs'(1)$,  incorporates terms linear in
$\lambda-1$ and so it provides a very good estimate of the SW
internal energy.

A more complete comparison between $y_{\sw}(r)$ and $y_{\shs}(r)$ is
provided by Figs.\ \ref{fig2}--\ref{fig4}. There it is shown that
the global shape of $y_{\sw}(r|\eta,T^*;\lambda)$ at fixed $\eta$
and $\tau_{\text{eff}}(T^*,\lambda)$ is only weakly influenced by
the value of $\lambda$. In particular, the precursors of the
singularities (delta-peaks and/or discontinuities)\cite{SG87,MF04a}
of $y_{\shs}(r)$ at $r=\sqrt{\frac{8}{3}}, \frac{5}{3}, \sqrt{3},
2,\ldots$ are clearly apparent in the SW fluids, especially for
$\lambda=1.01$. Apart from those local singularities, one can
observe that the SW cavity function for $\lambda\leq 1.02$ is
reasonably well described by the SHS one, while slight but visible
deviations are present in the case $\lambda=1.05$.

\begin{figure}
\includegraphics[width=\columnwidth]{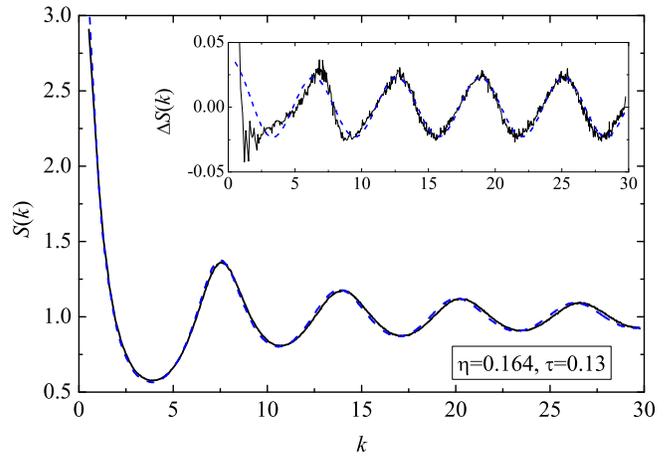} \caption{(Color online) Structure factor $S(k)$ at
$\eta=0.164$ and $\tau_{\text{eff}}=0.13$.  The solid line
corresponds to MC simulations for SHS by Miller and
Frenkel.\protect\cite{MF04a} The  dashed curve corresponds to our
simulations for an SW system with $\lambda=1.02$. The inset shows
the difference $\Delta S(k)\equiv S_{\sw}(k)-S_{\shs}(k)$ as
obtained from simulations (solid line) and as given by Eq.\
(\protect\ref{17}) (dashed line).}
\label{fig5}
\end{figure}

\begin{figure}
\includegraphics[width=\columnwidth]{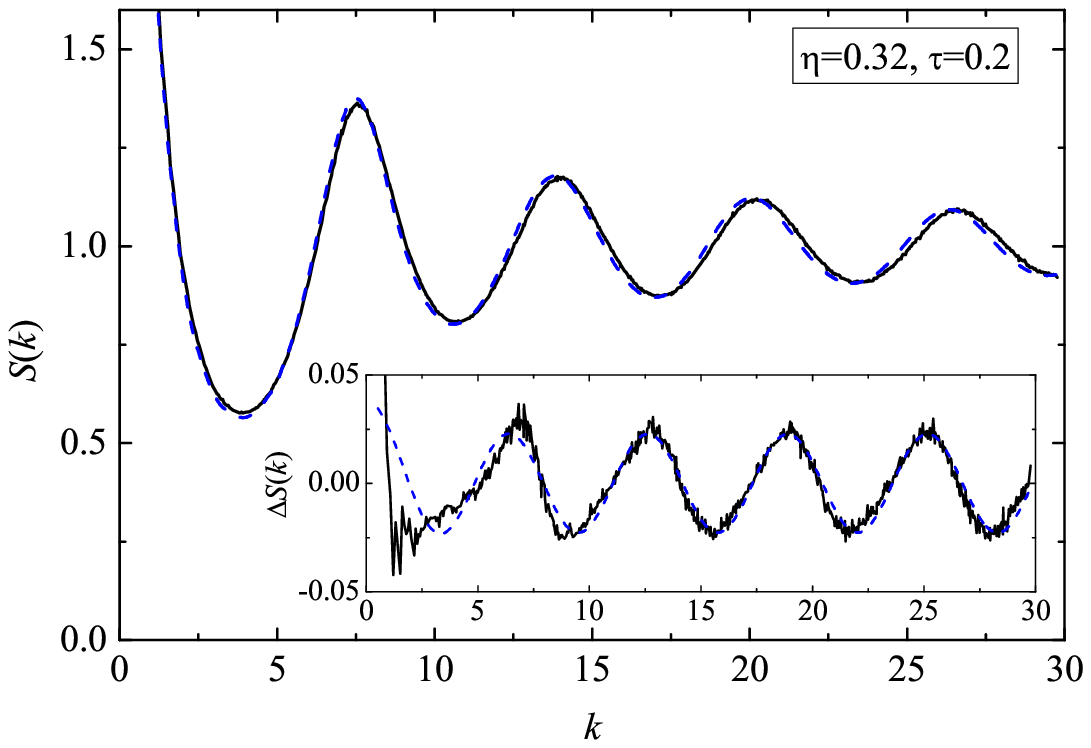} \caption{(Color online) Same as in Fig.\
\protect\ref{fig5} but for $\eta=0.32$ and $\tau_{\text{eff}}=0.2$.}
\label{fig6}
\end{figure}

\begin{figure}
\includegraphics[width=\columnwidth]{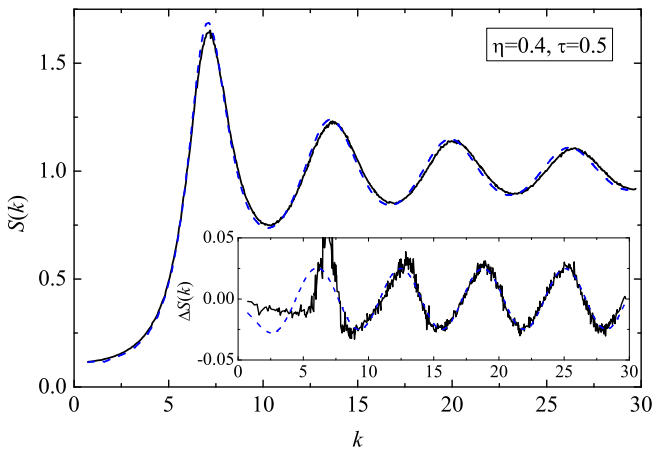} \caption{(Color online) Same as in Fig.\
\protect\ref{fig5} but for $\eta=0.4$ and $\tau_{\text{eff}}=0.5$.}
\label{fig7}
\end{figure}

It must be emphasized again that the good agreement between the SHS
and SW cavity functions does not imply the same in the case of the
radial distribution function inside the well ($1\leq r\leq
\lambda$). As a consequence, the structure factor, being related to
the Fourier transform of $h(r)$, can be expected to be affected by
the  difference between $g_{\shs}(r)$ and $g_{\sw}(r)$ near $r=1$.
To visualize this effect, we compare in Figs.\
\ref{fig5}--\ref{fig7} the SHS and SW structure factors, in the
latter case with $\lambda=1.02$. Although both functions are very
close each other in the three cases, some systematic differences are
apparent. In particular, the maxima and minima of $S_{\sw}(k)$ are
slightly more pronounced and shifted to the left than those of
$S_{\shs}(k)$. The difference $\Delta S(k)\equiv
S_{\sw}(k)-S_{\shs}(k)$ is plotted in the insets of Figs.\
\ref{fig5}--\ref{fig7}. Except for small $k$, it is observed that
the simulation data of $\Delta S(k)$ are remarkably well described
by the approximation (\ref{17}), where the simulation values for
$y_\shs(1)$ and  $y_\shs'(1)$ are used. This confirms that the
differences between $y_{\sw}(r)$ and $y_{\shs}(r)$, which are
neglected in Eq.\ (\ref{17}), have a practically irrelevant
influence on  the difference between $S_{\sw}(k)$ and $S_{\shs}(k)$.


\section{Comparison with theoretical predictions\label{sec4}}
The approximate mapping $y_{\sw}(r)\to y_{\shs}(r)$ can be exploited
to get theoretical predictions for the radial distribution function
$g_{\sw}(r)$ of short-range SW fluids from the availability of
analytical expressions for $g_{\shs}(r)$. In particular, we will use
Baxter's solution of the PY equation for SHS,\cite{B68} as well as
an extension of it proposed by two of us.\cite{YS93b} For the sake
of completeness, we will also consider an alternative theoretical
approach directly devised for SW fluids.\cite{YS94,AS01,LSAS03}
Before comparing with our MC simulations, the expressions stemming
from these three theories are briefly described.

\subsection{Percus--Yevick and rational-function approximations for
sticky hard spheres\label{subsec4.1}}

As is well known, Baxter was able to find the exact solution of the
PY theory for the case of SHS.\cite{B68} This solution was later
interpreted as the simplest case of a more general class of
approximations.\cite{YS93a,YS93b} Both approaches are constructed in
the Laplace space. Let us introduce the Laplace transform of $r
g(r)$:
\beq
G(s)\equiv\int_1^\infty \dd r\, e^{-r s} r g(r)=s
\frac{F(s)\ee^{-s}}{1+12\eta F(s)\ee^{-s}},
\label{4.1b}
\eeq
where the last equality defines the auxiliary function $F(s)$. In
real space,
\beq
g(r)=\frac{1}{r}\sum_{n=1}^\infty
(-12\eta)^{n-1}f_n(r-n)\Theta(r-n),
\label{4.2b}
\eeq
where $f_n(r)$ is the inverse Laplace transform of $s[F(s)]^n$ and
$\Theta(x)$ is Heaviside's step function. The auxiliary function
$F(s)$ must comply with some consistency conditions for small $s$
and for large $s$:\cite{YS93a,YS93b}
\beq
\frac{\ee^s}{F(s)}=-12\eta +s^3+\mathcal{O}(s^5),
\label{4.3}
\eeq
\beq
F(s)=y(1)s^{-1}\left(\frac{1}{12\tau}+s^{-1}\right)+\mathcal{O}(s^{-3}).
\label{4.4}
\eeq

A simple class of approximations consists of assuming a rational
form for $F(s)$:
\beq
F(s)=-\frac{1}{12\eta}\frac{1+\sum_{j=1}^m L_j
s^j}{1+\sum_{j=1}^{m+1} S_j s^j},
\label{4.5}
\eeq
with $m\geq 2$. The simplest case corresponds to $m=2$, the
coefficients  $L_1$, $L_2$, $S_1$, $S_2$, and $S_3$ being explicitly
determined \cite{YS93a} from the application of the physical
constraints (\ref{4.3}) and (\ref{4.4}). This  leads to the exact
solution of the PY equation for SHS. \cite{B68} One of the
shortcomings of this PY-SHS theory is that it yields inconsistent
equations of state via the energy and compressibility routes.

A more flexible approximation is obtained by setting $m=3$ in Eq.\
(\ref{4.5}) and  fixing the additional parameters $L_3$ and $S_4$ by
consistently imposing a given equation of state through the contact
value $y(1)$ and the isothermal susceptibility $\chi$.\cite{YS93b}
We will refer to the ansatz (\ref{4.5}) with $L_3$ and $S_4$
determined in this way as the rational-function approximation for
SHS (RFA-SHS). Given a compressibility factor $Z_\shs(\eta,\tau)$
for the SHS fluid, the isothermal compressibility and the excess
internal energy are obtained as\cite{YS93a}
\beq
\chi_\shs^{-1}=\frac{\partial}{\partial \eta}\left(\eta
Z_\shs\right),
\eeq
\beq
\frac{u_\shs}{\epsilon}=-\tau\frac{\partial}{\partial\tau}\int_0^\eta
\dd\eta'\frac{Z_\shs(\eta',\tau)-1}{\eta'}.
\label{4.6b}
\eeq
The contact value $y_\shs(1)$ consistent with $Z_\shs$ can then be
obtained from the internal energy through Eq.\ (\ref{uex2}). For the
compressibility factor $Z_\shs(\eta,\tau)$ we will use the empirical
form recently proposed by Miller and Frenkel.\cite{MF04b}

\subsection{Rational-function approximations for
square-well fluids} In the case of SW fluids with a hard-core
diameter $\sigma=1$ and range $\lambda$ one can still define the
auxiliary function $F(s)$ trough Eq.\ (\ref{4.1b}), so that Eqs.\
(\ref{4.2b}) and (\ref{4.3}) also hold in this
case.\cite{YS93a,YS93b} However, the constraint (\ref{4.4}) is now
replaced by
\beq
F(s)=g(1^+)s^{-2}+\mathcal{O}(s^{-3}).
\label{4.6}
\eeq

In the spirit of the ansatz (\ref{4.5}), the simplest
rational-function approximation for SW systems (RFA-SW) has the
form\cite{YS94,AS01}
\beq
F(s)=-\frac{1}{12\eta}\frac{1+A+K_1 s-(A+K_2
s)\ee^{-(\lambda-1)s}}{1+S_1 s+S_2 s^2+S_3 s^3}.
\label{4.7}
\eeq
Application of the condition (\ref{4.3}) allows one to express
$K_1$, $S_1$, $S_2$, and $S_3$ as linear functions of $A$ and $K_2$.
Next, $A$ is taken for simplicity as independent of density, so that
$A=\ee^{1/T^*}-1$. Finally, $K_2$ is obtained by enforcing the
continuity of $y(r)$ at $r=\lambda$.

It can be proved\cite{YS94} that in the SHS limit ($\lambda\to 1$,
$T^*\to 0$ with $\tau=\text{finite}$), one has $K_2(\lambda-1)\to
L_2$ and $K_1-K_2+A(\lambda-1)\to L_1$, so that Eq.\ (\ref{4.7})
reduces to Eq.\ (\ref{4.5}) with $m=2$, i.e., the PY-SHS solution.
An improved RFA-SW theory that would reduce to the RFA-SHS theory
would require the addition of new terms in the numerator and
denominator of Eq.\ (\ref{4.7}), but then extra constraints would be
needed to determine those new coefficients.

\subsection{Comparison}
\begin{figure}
\includegraphics[width=\columnwidth]{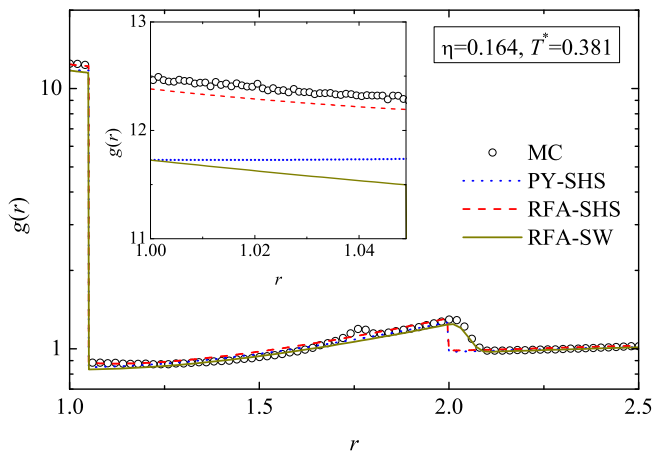} \caption{(Color online) Radial distribution
function of an SW fluid with $\lambda=1.05$ at $\eta=0.164$ and
$T^*=0.381$. The circles correspond to our MC simulations and the
solid lines represent the RFA-SW predictions. The results obtained
from the approximation (\protect\ref{3}) with
$\tau_{\text{eff}}=0.13$ by using the PY and RFA theories for the
SHS fluid are represented by the dotted and dashed lines,
respectively. The inset shows $g(r)$ inside the well. Outside that
region, the PY-SHS and RFA-SHS curves are practically
indistinguishable.}
\label{fig8}
\end{figure}

\begin{figure}
\includegraphics[width=\columnwidth]{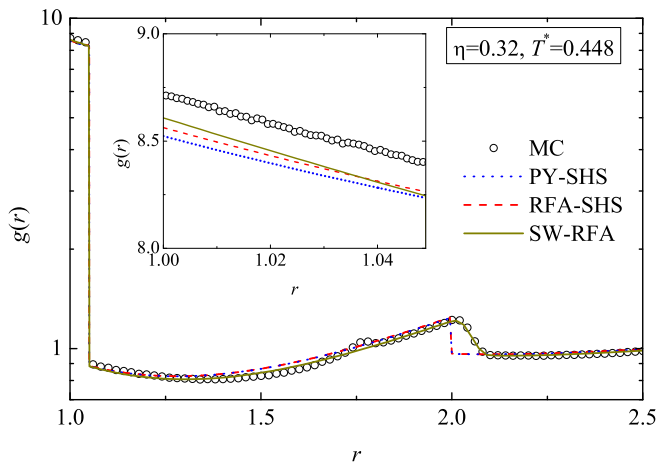} \caption{(Color online) Same as in Fig.\
\protect\ref{fig9} but for $\eta=0.32$, $T^*=0.448$ and
$\tau_{\text{eff}}=0.2$.}
\label{fig9}
\end{figure}

\begin{figure}
\includegraphics[width=\columnwidth]{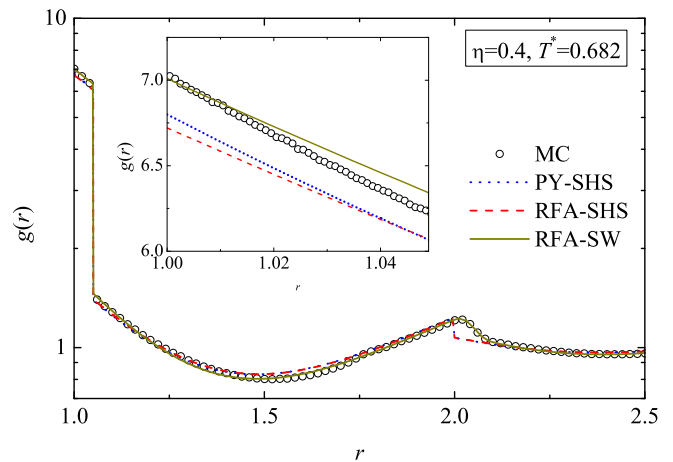} \caption{(Color online) Same as in Fig.\
\protect\ref{fig9} but for $\eta=0.4$, $T^*=0.682$ and
$\tau_{\text{eff}}=0.5$.}
\label{fig10}
\end{figure}

We are now in conditions of comparing the three theories against our
MC simulation data. We restrict ourselves to the case
$\lambda=1.05$. Since the radial distribution function $g(r)$ is a
more direct and intuitive quantity than the cavity function $y(r)$,
we present the plots in terms of $g(r)$ in Figs.\
\ref{fig8}--\ref{fig10}.
We observe a good general agreement between theories and simulations
for the three states. This is noteworthy since no fitting has been
employed. Only in the case of the RFA-SHS approach there exist two
free parameters ($L_3$ and $S_4$) which are determined by enforcing
thermodynamic consistency with an empirical equation of state for
SHS fluids,\cite{MF04b} as explained above. Despite the good
theoretical behaviour, none of the theories predict the distortion
around $r=1.77$, which is the precursor of the singularity of the
exact $g_{\shs}$ at $r=\sqrt{3}\approx 1.73$. Obviously, the PY-SHS
and RFA-SHS approximations transfer to $g_{\sw}$ the discontinuity
of $g_{\shs}$ at $r=2$, while the RFA-SW approximation accurately
captures the rapid but continuous decay of $g_{\shs}$ near $r=2$. In
general, both SHS theories yield results which are practically
indistinguishable, especially for the two highest temperatures
($T^*=0.448$ and $T^*=0.682$). Only inside the attractive well
($1\leq r\leq 1.05$) can one distinguish the PY-SHS and  RFA-SHS
predictions, as shown in the insets. This region is especially
important for SW fluids since it is directly related to the
coordination number and the excess internal energy. We observe that
the best performance in the region $1\leq r\leq 1.05$ is done by the
RFA-SHS theory at the lowest temperature ($T^*=0.381$) and by the
RFA-SW theory at the other two temperatures ($T^*=0.448$ and
$T^*=0.682$). In this respect, it must be borne in mind that the
choice (\ref{13}) of the effective stickiness parameter is more
appropriate for temperatures such that $\tau_{\text{eff}}\approx
0.1$ than for higher temperatures.

\section{Mixtures of SW fluids\label{sec5}}
Let us now consider the case of multicomponent systems of particles
interacting via SW potentials:
\beq
\varphi_{ij}^\sw(r)=
\begin{cases}
\infty,& r<\sigma_{ij},\\
-\epsilon_{ij},& \sigma_{ij}<r<\lambda_{ij}\sigma_{ij},\\
0,& r>\lambda_{ij}\sigma_{ij}.
\end{cases}
\label{B1}
\eeq
In the one-component case, there are three parameters characterizing
the interaction ($\sigma$, $\epsilon$, and $\lambda$) and two
parameters defining the thermodynamic state ($\rho$ and $T$).
Without loss of generality, we can use $\sigma$ and $\epsilon$ to
fix the units of distance and energy, respectively, so that only
three independent dimensionless parameters remain ($\lambda$,
$\eta$, and $T^*$). In the case of mixtures, however, the  number of
parameters is in general much higher. Restricting ourselves to the
binary case, we have  three thermodynamic quantities (the total
density $\rho$, the mole fraction $x_1$ of one of the species, and
the temperature $T$) plus nine interaction parameters (the diameters
$\sigma_{ij}$, the well depths $\epsilon_{ij}$, and the relative
ranges $\lambda_{ij}$). One of the diameters can be used to define
the length unit and one of the depths can be used to define the
energy unit, so that the number of independent parameters defining
the problem is ten in general.\cite{note1}

The SW potentials (\ref{B1}) become SHS potentials if
$\lambda_{ij}\to 1$ and $\epsilon_{ij}\to\infty$ by keeping constant
the parameters
\beq
\tau_{ij}^{\text{eff}}=\frac{1}{12(\ee^{\epsilon_{ij}/k_BT}-1)(\lambda_{ij}-1)}.
\label{B3}
\eeq
Now, each pair $(\epsilon_{ij},\lambda_{ij})$ gives birth to a
 parameter $\tau_{ij}^{\text{eff}}$ and so the general number of parameters
 characterizing the binary mixture reduces from ten to seven.
Note that,  although the term $-1$ appearing in the denominator of
Eq.\ (\ref{B3}) can be neglected in the SHS limit, we have kept it
so that Eq.\ (\ref{B3}) defines the generalization to mixtures of
the effective stickiness parameter defined by Eq.\ (\ref{13}). The
exact solution of the PY equation for SHS mixtures is
known\cite{PS75} in the \textit{additive} case
$\sigma_{ij}=(\sigma_{ii}+\sigma_{jj})/2$.  Systems of SHS mixtures
have been considered by several
authors.\cite{BT79,JBH91,TKR02,GG02,FGG05,J06} An extension of the
RFA described in Subsec.\ \ref{subsec4.1} to SHS mixtures is also
available.\cite{SYH98}

In agreement with the philosophy behind the approximation (\ref{3})
for the one-component case, we can expect that, for narrow SW
potentials,
\begin{widetext}
\beq
y_{ij}^\sw(r|\rho,x_1,T,\{\sigma_{k\ell}\},\{\epsilon_{k\ell}\},\{\lambda_{k\ell}\})\approx
y_{ij}^\shs
(r|\rho,x_1,\{\sigma_{k\ell}\},\{\tau_{k\ell}^{\text{eff}}\}),
\label{5.5}
\eeq
\end{widetext}
where $y_{ij}(r)\equiv \ee^{\varphi_{ij}(r)/k_BT}g_{ij}(r)$.

Given the high dimensionality of the parameter space for binary SW
mixtures, we restrict ourselves to the two classes of mixtures shown
in Table \ref{table2}. Class I corresponds  to equimolar and
symmetric mixtures in which the self interactions 1-1 and 2-2 are of
HS type, while the cross interaction 1-2 is of SW type. Therefore,
$\tau_{11}^{\text{eff}}=\tau_{22}^{\text{eff}}=\infty$ and
$\tau_{12}^{\text{eff}}=\tau_{\text{eff}}$. This class of mixtures
can be used to describe adsorption phenomena.\cite{BT79} Class II is
representative of asymmetric hard-core diameters, but with common
well depths and relative well widths, so that
$\tau_{ij}^{\text{eff}}=\tau_{\text{eff}}$. It only remains to fix
for each class the values of the range $\lambda$, the reduced
temperature $T^*=k_BT/\epsilon$ and the reduced density
$\rho\sigma^3$ or, equivalently, the total packing fraction
$\eta=\pi\rho(x_1\sigma_{11}^3+x_2\sigma_{22}^3)/6$. The values
considered in this paper for those quantities are shown in Table
\ref{table3}. This yields a total of 12 different mixtures
simulated. For each $\lambda$, the temperature is chosen so that
$\tau_{\text{eff}}=0.2$ and 0.5, in analogy with some of the cases
considered in Table \ref{table1}.
\begin{table}
\begin{ruledtabular}
\begin{tabular}{cccccccccccc}
Class&   $\sigma_{11}$& $\sigma_{22}$&   $\sigma_{12}$&
$\lambda_{11}$&   $\lambda_{22}$& $\lambda_{12}$&   $\epsilon_{11}$&
$\epsilon_{22}$&
$\epsilon_{12}$&$x_1$\\
\hline
I&1&1&1&1&1&$\lambda$&0&0&$\epsilon$&$\frac{1}{2}$\\
II&1&3&2&$\lambda$&$\lambda$&$\lambda$&$\epsilon$&$\epsilon$&$\epsilon$&$\frac{15}{16}$
\end{tabular}
 \caption{Classes of SW mixtures considered in the simulations.}
\label{table2}
\end{ruledtabular}
\end{table}
\begin{table}
\begin{ruledtabular}
\begin{tabular}{cccc}
$\eta$& $\tau_{\text{eff}}$&$\lambda$& $T^*$\\
\hline
0.4&0.2&1.05&0.448\\
&&1.02&0.324\\
&&1.01&0.266\\
&0.5&1.05&0.682\\
&&1.02&0.448\\
&&1.01&0.348
\end{tabular}
 \caption{Values of $\eta$, $\lambda$ and $T^*$ considered in the simulations.}
\label{table3}
\end{ruledtabular}
\end{table}

Now we compare, for each class and each value of
$\tau_{\text{eff}}$, the cavity functions $y_{ij}(r)$ obtained from
our MC simulations. Given the scarcity  of simulations for SHS
mixtures,\cite{note2} we use the analytical solution of the PY
closure.\cite{PS75} We cannot use the expressions provided by the
RFA method\cite{SYH98} because the latter relies upon the knowledge
of the equation of state for SHS mixtures. While an accurate
empirical equation of state exists for one-component SHS
fluids,\cite{MF04b} its extension to mixtures is not known.

\begin{figure}
\includegraphics[width=\columnwidth]{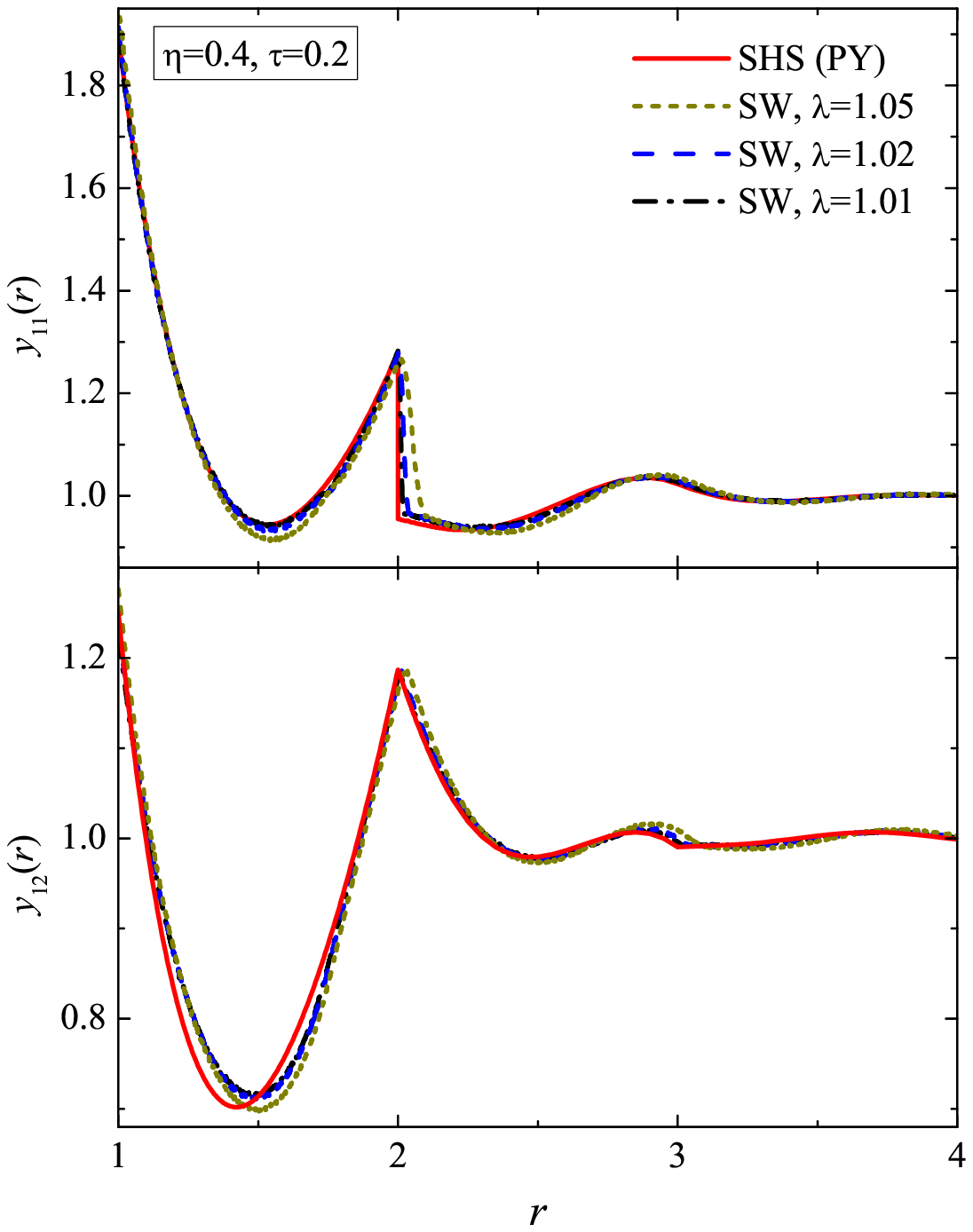} \caption{(Color online) Cavity functions at
$\eta=0.4$ and $\tau_{\text{eff}}=0.2$ for a binary mixture of class
I (see Table \protect\ref{table2}). The solid lines correspond to
the analytical solution of the PY theory for SHS mixtures. The
dotted, dashed, and dash-dotted curves correspond to our simulations
for SW systems with $\lambda=1.05$, $1.02$, and $1.01$,
respectively.}
\label{fig11}
\end{figure}
\begin{figure}
\includegraphics[width=\columnwidth]{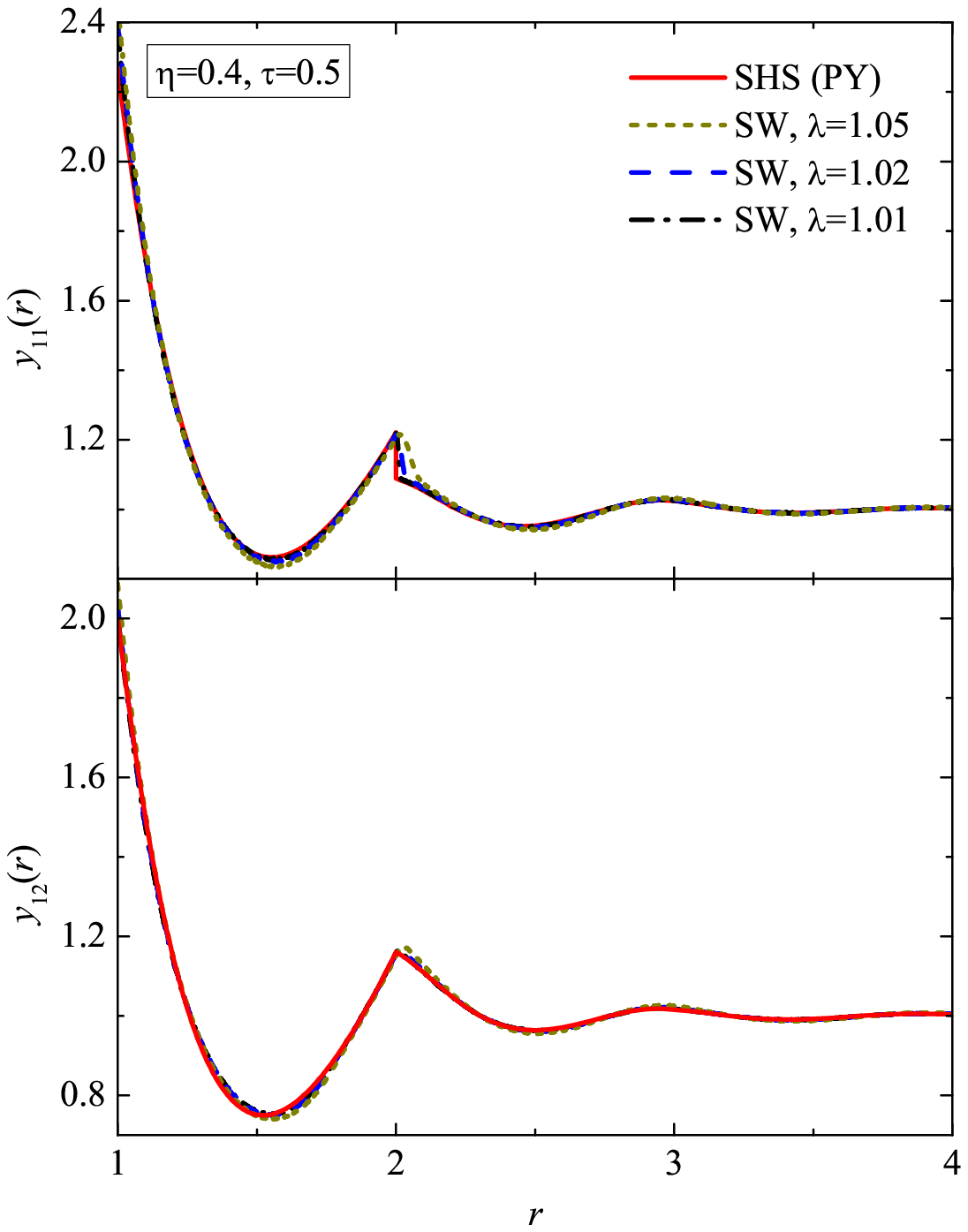} \caption{(Color online) Same as in Fig.\
\protect\ref{fig11} but for $\eta=0.4$ and $\tau_{\text{eff}}=0.5$.}
\label{fig12}
\end{figure}
\begin{figure}
\includegraphics[width=\columnwidth]{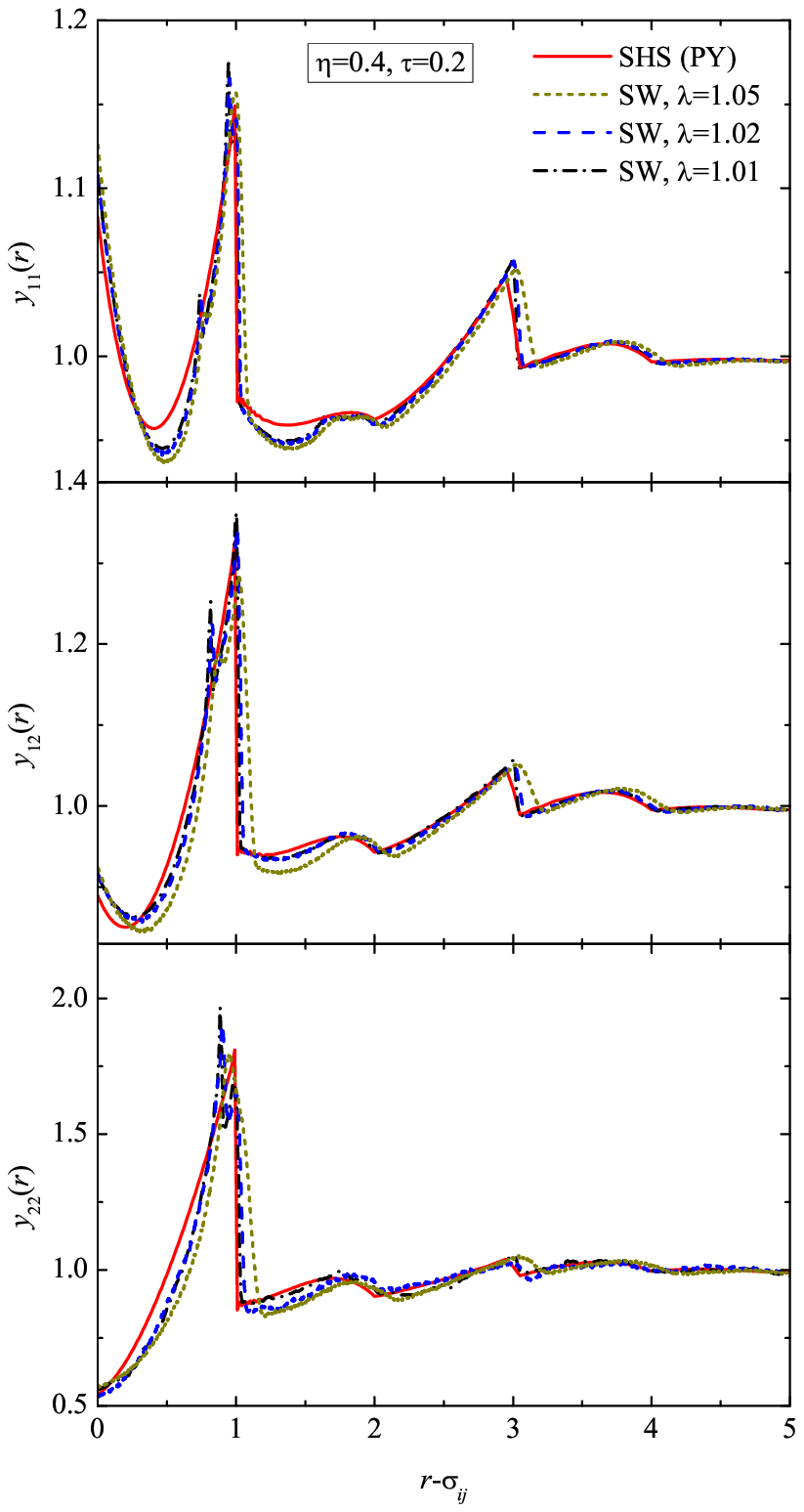} \caption{(Color online) Cavity functions at
$\eta=0.4$ and $\tau_{\text{eff}}=0.2$ for a binary mixture of class
II (see Table \protect\ref{table2}). The solid lines correspond to
the analytical solution of the PY theory for SHS mixtures. The
dotted, dashed, and dash-dotted curves correspond to our simulations
for SW systems with $\lambda=1.05$, $1.02$, and $1.01$,
respectively.}
\label{fig13}
\end{figure}
\begin{figure}
\includegraphics[width=\columnwidth]{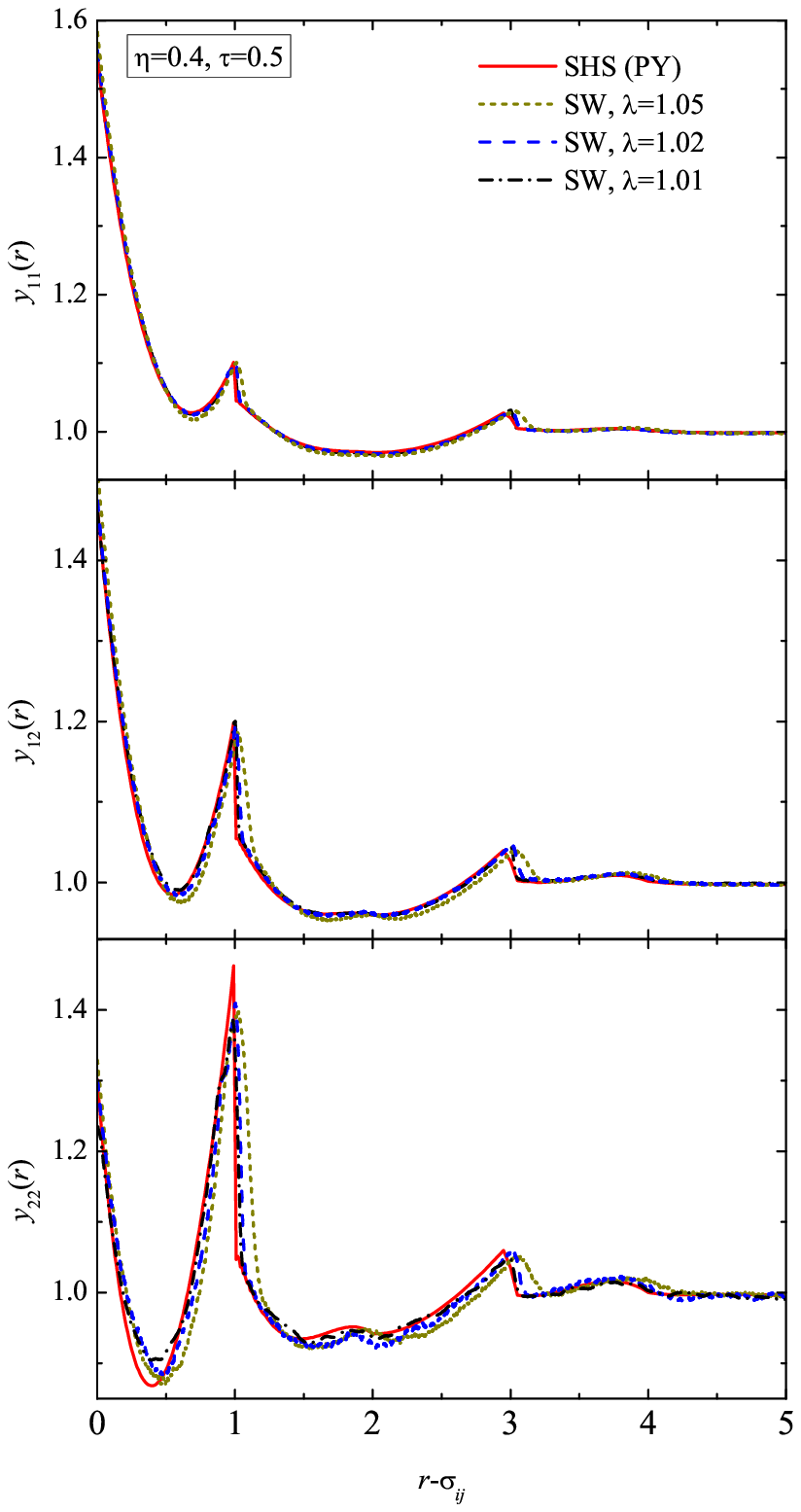} \caption{(Color online) Same as in Fig.\
\protect\ref{fig13} but for $\eta=0.4$ and $\tau_{\text{eff}}=0.5$.}
\label{fig14}
\end{figure}

The two states corresponding to mixtures of class I are considered
in Figs.\ \ref{fig11} and \ref{fig12}. We observe that the range
$\lambda=1.05$ is again not small enough to produce a good overlap
with the cases $\lambda=1.02$ and $\lambda=1.01$. These two latter
SW mixtures can be expected to be close to the  true SHS mixture. In
this respect, the deviations of our SW simulations with
$\lambda=1.02$ and $\lambda=1.01$ from the SHS-PY theory reveal the
limitations of the latter.\cite{J06} Figures \ref{fig11} and
\ref{fig12} show that the PY theory describes better the pair
correlation function $y_{11}=y_{22}$ of the particles interacting
via the HS potential than the pair correlation function $y_{12}$ of
the particles feeling a short-range attraction. Accordingly, the
agreement improves when the temperature increases (from
$\tau_{\text{eff}}=0.2$ to $\tau_{\text{eff}}=0.5$) and hence the
stickiness decreases.

The results corresponding to the asymmetric mixtures of class II are
plotted in Figs.\ \ref{fig13} and \ref{fig14}. The convergence
SW$\to$SHS is similar to that in the symmetric cases since the
deviations between the curves corresponding to the ranges
$\lambda=1.02$ and $\lambda=1.01$ are again small. However, the
performance of the PY theory is worse in Figs.\ \ref{fig13} and
\ref{fig14} than in Figs.\ \ref{fig11} and \ref{fig12}. This could
have been expected by the following arguments. On the one hand, the
presence of stickiness is larger in class II than in class I. On the
other hand, it is known that the PY theory is more accurate for
one-component HS fluids (class I in the limit $\tau\to\infty$) than
for HS mixtures (class II in the limit
$\tau\to\infty$).\cite{MMYSH02}

\section{Summary and concluding remarks\label{sec6}}

In this paper we have investigated the possibility of representing
the structural and thermodynamic properties of short-range SW fluids
by those of SHS fluids.  It is sometimes  assumed in the literature
on colloidal suspensions that a range $\lambda \leq 1.1$ is small
enough to replace the more realistic SW model by the simpler SHS
model. Moreover, the mapping SW$\to$SHS is usually assumed to hold
at the level of the structure factor, i.e., $S_\sw(k)\approx
S_\shs(k)$. This in turn implies $g_\sw(r)\approx g_\shs(r)$, what
misses the fact that $g_\sw(r)$ has a jump discontinuity at
$r=\lambda$ which becomes a delta singularity of $g_\shs(r)$ at
$r=1^+$.

Here, however, the ansatz $S_\sw(k)\approx S_\shs(k)$ or,
equivalently, $g_\sw(r)\approx g_\shs(r)$ has been replaced by
$y_\sw(r)\approx y_\shs(r)$. This still leaves open the problem of
finding the effective packing fraction $\eta_{\text{eff}}$ and the
effective stickiness parameter $\tau_{\text{eff}}$ of the SHS fluid
that best mimics the physical properties of an SW fluid of range
$\lambda$ at a given state $(\eta,T^*)$. For simplicity, we have
chosen $\eta_{\text{eff}}$ to be independent of $T^*$ and
$\tau_{\text{eff}}$ to be independent of $\eta$. Next, the
constraint of recovering the HS fluid in the limit $T^*\to\infty$
leads to $\eta_{\text{eff}}=\eta$ in a natural way. Since
$\tau_{\text{eff}}$ is assumed to be independent of $\eta$, we have
taken advantage of the exact knowledge of $y_\sw(r)$ and $y_\shs(r)$
to first order in density as a guide to choose $\tau_{\text{eff}}$.
Considering several possibilities of the form (\ref{6}), we have
found that the optimal choice for temperatures such that
$\tau_{\text{eff}}\approx 0.1$ is provided by Eq.\ (\ref{13}), as
illustrated by Fig.\ \ref{fig1}. This choice differs from the
conventional one,\cite{SG87,RR89,TB93,VL00} based on the equality of
the SW and SHS second virial coefficients, i.e.,
$B_2^{\sw}(T^*;\lambda)=B_2^{\shs}(\tau_{\text{eff}}(T^*,\lambda))$.

The condition $B_2^{\sw}=B_2^{\shs}$ is directly related to the
assumption that the thermodynamic properties of the SW and SHS
fluids are the same, at least for low densities. Our approach
differs from this one. We have assumed that the \textit{cavity
function} is approximately the same in both systems and from this
ansatz we have obtained expressions for  the structure factor, the
internal energy, the isothermal compressibility, and the pressure of
the SW fluid in terms of quantities related to the SHS fluid. Those
approximate relations have been checked by simulation data in the
case of the structure factor (see Figs.\ \ref{fig5}--\ref{fig7}) and
the internal energy (see Table \ref{table1}).

We have performed MC simulations of one-component SW fluids for
$\lambda=1.05$, 1.02, and 1.01. For each case, we have considered
the densities and temperatures indicated in Table \ref{table1}. They
correspond to the same values of the packing fraction and effective
stickiness as considered by Miller and Frenkel in their simulations
of SHS fluids.\cite{MF04a} The comparisons (see Figs.\
\ref{fig2}--\ref{fig4}) show that the SW cavity functions with
$\lambda=1.05$ present small but visible differences with respect to
the SHS ones. On the other hand, the SW curves corresponding to
$\lambda=1.02$ and 1.01 have almost collapsed to the  SHS curves,
exhibiting the precursors of the singularities of $y_\shs(r)$ at
$r=\sqrt{\frac{8}{3}}, \frac{5}{3}, \sqrt{3}, 2,\ldots$

While the PY equation has an exact solution for SHS, its solution
for SW is not known exactly. However, we have exploited the
knowledge of the PY-SHS solution,\cite{B68} as well as of the
rational-function approximation (RFA) for SHS,\cite{YS93b}  to
estimate the radial distribution function $g_\sw(r)$ based on the
ansatz $y_\shs(r)\approx y_\sw(r)$. The results for $\lambda=1.05$
show a good general agreement with the MC data (see Figs.\
\ref{fig8}--\ref{fig10}), except near $r=2$, where $g_\sw(r)$ decays
rapidly but does not present the jump discontinuity of $g_\shs(r)$.
{}From that point of view, a better agreement is provided by the
RFA-SW theory.\cite{YS94}

In order to offer a wider perspective, we have also considered two
classes of binary mixtures (see Table \ref{table2}). Class I defines
a  symmetric equimolar mixture where the interactions among
particles of the same species are of HS type, while those among
particles of different species are of SW type. Mixtures of class II
are asymmetric with a hard-core ratio 1:3 but with an attractive
interaction of common depth and (relative) width. Comparison between
our simulation data for SW mixtures  and the theoretical PY-SHS
predictions\cite{PS75} shows that the latter behaves better for
mixtures of class I (see Figs.\ \ref{fig11} and \ref{fig12}) than
for mixtures of class II (see Figs.\ \ref{fig13} and \ref{fig14}).
In each case, the agreement improves  as the temperature increases
(and hence the stickiness decreases). Apart from this comparison
with the PY-SHS theory, our MC simulations show again that the range
$\lambda=1.05$ is not small enough to get a good collapse with the
cases $\lambda=1.02$ and 1.01.

One of the open avenues not fully explored in this paper refers to
the analysis of the equation of state for short-range SW fluids
obtained from the empirical equation of state for SHS fluids
recently proposed by Miller and Frenkel.\cite{MF04b} Equations
(\ref{14}), (\ref{17bis}), and (\ref{ZSWSHS}) provide three
(approximate) alternative routes to the SW equation of state. They
require the knowledge of $y_\shs(1)$, $y'_\shs(1)$, and
$y''_\shs(1)$. The two former quantities can be obtained from the
SHS equation of state through Eqs.\ (\ref{ZShs}) and (\ref{uex2}),
while $y''_\shs(1)$ can be obtained from the RFA-SHS
theory.\cite{YS93b} We plan to undertake this study in the near
future.

\acknowledgments

We are very grateful to Dr.\ M. A. Miller  for providing us with
simulation data for sticky hard spheres. Al.M. thanks the Junta de
Extremadura for supporting his stay at the University of Extremadura
in the period October--December 2005, when most of the computational
 work was done. His research has been
 partially supported by the Ministry of Education, Youth, and Sports of
the Czech Republic under the project LC 512 and by the Grant Agency
of the Czech Republic under project No. 203/06/P432. The research of
S.B.Y. and A.S. has been supported by the Ministerio de Educaci\'on
y Ciencia (Spain) through Grant No. FIS2004-01399 (partially
financed by FEDER funds) and by the European Community's Human
Potential Programme under contract HPRN-CT-2002-00307, DYGLAGEMEM.

\appendix*

\section{The cavity function to first order in density\label{app}}
To first order in density, the cavity function of the SW fluid is
given by Eq.\ (\ref{A1}). The coefficient $y_\sw^{(1)}$
is\cite{BH67}
\begin{widetext}
\beq
y_{\sw}^{(1)}(r|T^*;\lambda)=A(r)+(\ee^{1/T^*}-1)B_{\sw}(r|\lambda)+(\ee^{1/T^*}-1)^2C_{\sw}(r|\lambda),
\label{A2}
\eeq
where (for $\lambda\leq 3$)
\beq
A(r)=
\begin{cases}
8-6r+\frac{1}{2}r^3,&r< 2,\\
0,&r>2,
\end{cases}
\label{A3}
\eeq
\beq
B_{\sw}(r|\lambda)=
\begin{cases}
-12r+r^3,&r< \lambda-1,\\
3(\lambda^2-1)^2
r^{-1}-8(\lambda^3-1)+6(\lambda^2-1)r,&\lambda-1<r< 2,\\
3(\lambda^2-1)^2r^{-1}-8(\lambda^3+1)+6(\lambda^2+1)r-r^3,&2<r< \lambda+1,\\
0,&r>\lambda+1,
\end{cases}
\label{A4}
\eeq
\beq
C_{\sw}(r|\lambda)=
\begin{cases}
8(\lambda^3-1)-6(\lambda^2+1)r+r^3,&r< \lambda-1,\\
3(\lambda^2-1)^2
r^{-1},&\lambda-1<r<2,\\
3(\lambda^2-1)^2r^{-1}-8+6r-\frac{1}{2}r^3,&2<r<\lambda+1,\\
8\lambda^3-6\lambda^2 r+\frac{1}{2}r^3,&\lambda+1<r<
2\lambda,\\
0,&r>2\lambda.
\end{cases}
\label{A5}
\eeq
\end{widetext}

Taking the SHS limit, Eq.\ (\ref{A2}) reduces to
\beq
y_{\shs}^{(1)}(r|\tau)=A(r)+\frac{1}{12\tau}B_{\shs}(r)+\frac{1}{(12\tau)^2}C_{\shs}(r),
\label{A7}
\eeq
where
\beq
B_{\shs}(r)=
\begin{cases}
12(r-2),&0<r< 2,\\
0,&r>2,
\end{cases}
\label{A8}
\eeq
\beq
C_{\text{SHS}}(r)=
\begin{cases}
12r^{-1},&0<r< 2,\\
0,&r>2.
\end{cases}
\label{A9}
\eeq


\begin{thebibliography}{99}
\bibitem{note1}
This number reduces to eight if we assume the Lorentz-Berthelot
rule, according to which $\sigma_{12}=(\sigma_{11}+\sigma_{22})/2$
and $\epsilon_{12}=\sqrt{\epsilon_{11}\epsilon_{22}}$.
\bibitem{note2}
We are not aware of simulations of SHS mixtures apart from the ones
recently reported by Jamnik in Ref.\ \protect\onlinecite{J06}, which
have been published after our SW simulations were finished.
\end{thebibliography}
\end{document}